\documentclass[prl,twocolumn,showpacs,preprintnumbers,amsmath,amssymb]{revtex4} 

\usepackage[dvips]{graphicx}
\usepackage{subfigure}
\usepackage{rotating}
\usepackage{hyperref}
\usepackage{color}
\DeclareGraphicsExtensions{.eps}

\def\eq{\begin{equation}}
\def\qe{\end{equation}}
\def\eqa{\begin{eqnarray}}
\def\qea{\end{eqnarray}}

\newcommand{\newc}{\newcommand}

\newc{\new}[1]{{\color{red}#1}}

\newc{\herwig}{\texttt{Herwig++\,}}
\newc{\herwigv}{\texttt{Herwig++ v2.4.2\,}}
\newc{\DJpsiFDC}{\texttt{DJpsiFDC\,}}

\newc{\scale}{0.75}

\newc{\ifb}{\textrm{fb}^{-1}}

\newc{\fb}{{\rm fb}}
\newc{\pb}{{\rm pb}}
\newc{\nb}{{\rm nb}}
\newc{\mb}{{\rm mb}}

\newc{\sigeff}{\sigma_{\rm eff}}
\newc{\sigDPS}{\sigma_{\rm DPS}}
\newc{\sigSPS}{\sigma_{\rm SPS}}

\newc{\mstw}{\texttt{MSTW 2008 NLO\,}}

\newc{\Jpsi}{J/\psi}
\newc{\jpsi}{\Jpsi}
\newc{\mup}{\mu^+}
\newc{\mum}{\mu^-}

\newc{\mmumu}{m_{\mu^+\mu^-}}
\newc{\pt}{p_{T}}
\newc{\mpt}{\langle p_{T}\rangle}
\newc{\dR}{\Delta R}
\newc{\dphi}{\Delta \phi}
\newc{\deta}{\Delta \eta}
\newc{\rap}{y}
\newc{\drap}{\Delta\rap}
\newc{\drapmin}{|\Delta\rap|_{\rm min}}

\begin{document}

\preprint{CAVENDISH-HEP-2011-07, TTK-11-16}

\title{Pair production of $\jpsi$ as a probe of double parton
  scattering at LHCb}

\author{C.~H.~Kom$^a$} 
\author{A.~Kulesza$^b$} 
\author{W.~J.~Stirling$^a$} 
\affiliation{$^a$Cavendish Laboratory, J.J. Thomson Avenue, Cambridge CB3
  0HE, United Kingdom}
\affiliation{$^b$Institute for Theoretical Particle Physics and Cosmology,
  RWTH Aachen University D-52056 Aachen, Germany}


\begin{abstract}
We argue that the recent LHCb observation of $\jpsi$--pair production
indicates a significant contribution from double parton scattering, in
addition to the standard single parton scattering component.  We
propose a method to measure the double parton scattering at LHCb using
leptonic final states from the decay of two prompt $\jpsi$ mesons.

\end{abstract}


\maketitle

{\bf Introduction.} The Large Hadron Collider (LHC) provides a unique
environment for precise measurements of hitherto poorly understood
phenomena. Since the flux of incoming partons increases with the
collision energy, there is a high probability at the LHC of
multiparton scattering, i.e. scattering of more than one pair of
partons in the same hadron--hadron collision. The parton--parton
correlations and distributions of multiple partons within a proton are
difficult to address within the framework of perturbative
QCD. Therefore detailed experimental studies of multi--parton
interactions are of great importance. In particular, it is widely
expected that measurements of double parton scattering (DPS) processes
with final states carrying relatively large transverse momentum
($\pt$) will provide relevant information on the nature of multiple
scattering. Probing DPS processes using leptonic final states has been
discussed in~\cite{leptonic}.  In this Letter, we discuss how
observing four--muon final states from pair production of $\jpsi$
could provide additional experimental input.

Taking advantage of high jet production rate at hadron colliders, DPS
searches have been performed by the AFS~\cite{AFS}, UA2~\cite{UA2},
CDF~\cite{CDF} and D0~\cite{D0} Collaborations in the four--jets (4$j$)
and $\gamma+3j$ channels. At the LHC, the pair production of muons
from single $\jpsi$ production benefits from a large cross section.
This implies a significant DPS production rate of two $\jpsi$
particles, which can subsequently decay into four muons, leading to
much cleaner signals compared with those in the jet--based studies.
In addition, the measurement of these $\jpsi$ pairs provides
complementary information on parton--parton correlations, as the hard
processes can be initiated by partons different from those leading to
4$j$ and $\gamma+3j$ events.

A pair of $\jpsi$ mesons can also be produced in a \emph{single}
parton scattering (SPS) process, e.g. $gg\to\jpsi\jpsi$.  Studies of
these SPS processes are expected to provide important insights for
improving the theoretical description of single quarkonium production.
Understanding the DPS contribution is thus an important task in this
context.  However, here we are primarily interested in the DPS as a
signal process and will regard the SPS production as an irreducible
background.

\begin{figure}
  \includegraphics[scale=\scale]{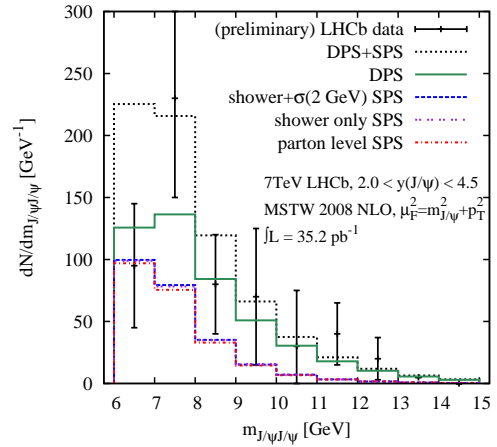}
  \caption{Invariant mass distribution of the $\jpsi$ pair
    ($m_{\jpsi\jpsi}$). The LHCb data are read off the plot
    in~\cite{LHCb-CONFNote}.  From top to bottom, the histograms are
    the DPS+SPS, DPS only and three (practically degenerate) SPS
    predictions including different radiation effects.
    \label{fig:res2M}}
\end{figure}

{\bf Invariant mass distribution of $\jpsi$--pairs at LHCb.}  The low
$\pt$ muon trigger at LHCb provides an ideal laboratory for studying
four--muon final states from the decay of two $\jpsi$ mesons. The LHCb
Collaboration has recently reported a first
measurement~\cite{LHCb-CONFNote} of this process, and compared the
two--$\jpsi$ invariant mass ($m_{\jpsi \jpsi}$) distribution with the
theoretical prediction of~\cite{Berezhnoy:2011xy} from direct SPS
production of two $\jpsi$ particles using a leading--order (LO)
color--singlet result first derived in~\cite{Qiao:2002rh}.  In
Fig.~\ref{fig:res2M}, we show a comparison between the data and an SPS
prediction for the process $gg\to\jpsi\jpsi$, which we implement in
the Monte Carlo event generator \herwigv\cite{Bahr:2008pv}.  This
process is also implemented in the event generator
\DJpsiFDC\cite{Qiao:2010kn}.  We neglect the known LO color--octet
contributions~\cite{Li:2009ug,Ko:2010xy} as they are predicted to be
negligible at the LHC~\cite{Ko:2010xy}.  In the calculations, the wave
function at the origin takes the value $|R(0)|^2=0.92$ ${\rm GeV}^3$.
We use the MSTW 2008 NLO PDFs~\cite{Martin:2009iq} and set the
renormalization ($\mu_R$) and factorization scales ($\mu_F$) equal to
$\sqrt{m_{\jpsi}^2+\pt^2}$, where $m_{\jpsi}=3.096$ GeV is the
physical $\jpsi$ mass.  The $\jpsi$ mesons are produced on--shell and
decay isotropically into $\mup\mum$ pairs with the branching ratio
$BR(\jpsi\to\mup\mum)= 0.05935$. This approximation, ignoring possible
$\jpsi$ polarization effects, is justified if one assumes small values
of $\jpsi$ transverse momentum~\cite{Abulencia:2007us}.  In our
studies we require, in correspondence with the experimental analysis,
that the $\jpsi$ mesons have rapidities $2 < y_{\jpsi} < 4.5$ and
transverse momenta $p_{T,\jpsi}<10$ GeV.

The framework of \herwig~allows us to include effects of the initial
state radiation (ISR) and intrinsic $\pt$ of initial state partons.
We set the root mean square intrinsic $\pt$ of a Gaussian model in
$\herwig$ to 2 GeV.  In the figures, they are referred to as
``shower'' and ``intrinsic'' respectively.  As can be seen in
Fig.~\ref{fig:res2M}, the impact of these effects on the predicted
$m_{\jpsi\jpsi}$ distribution is negligible.

Fig.~\ref{fig:res2M} shows that the $m_{\jpsi \jpsi}$ shape of the
standard SPS contribution does not match the data very well.  In
particular, it peaks at too low a $m_{\jpsi\jpsi}$ value.  The same
conclusion has also been derived in~\cite{LHCb-CONFNote} for the
theoretical predictions of~\cite{Berezhnoy:2011xy}. In addition to the
SPS contribution, in Fig.~\ref{fig:res2M} we also show the invariant
mass distributions for the DPS process, which will be discussed
shortly.  The DPS contribution has a broader shape and peaks at higher
values of $m_{\jpsi \jpsi}$, leading to much better agreement between
the combined SPS+DPS prediction and data than with the SPS
contribution alone.  The LHCb data could therefore indicate the
presence of a significant DPS contribution to the double $\jpsi$
production process.

Our predictions for the DPS double $\jpsi$ production are obtained
using an approximation in which the DPS cross section ($\sigDPS^{2 \,
  \jpsi}$) factorizes into a product of two SPS cross sections
($\sigSPS^{\jpsi}$):
\eq
\label{eq:DPS}
d \sigDPS^{2 \,\jpsi} = 
\frac{d \sigSPS^{\jpsi} \, d \sigSPS^{\jpsi}}{2 \sigeff},
\qe
where
\eq
\label{eq:SPS}
d \sigSPS^{\jpsi} = \sum_{a,b} f_a (x_a,\mu_F) f_b (x_b,\mu_F)\, d \hat{\sigma}_{\rm SPS}^{\jpsi} \, d x_a dx_b \,,
\qe
is the SPS cross section of a \emph{single} $\jpsi$, and
\eq
\label{eq:partSPS}
d \hat{\sigma}_{\rm SPS}^{\jpsi} = \sum_{a,b} \, \frac{1}{2 \hat{s}}
\, \overline{| M_{a b \to \jpsi +X} |^2} \, d{\rm PS}_{\jpsi+X}\, \qe
is the corresponding parton level process, with $\hat{s}$ the partonic
centre--of--mass energy.  The approximation assumes factorization
between the longitudinal and the transverse components of the
generalized double parton distributions and the assumption of no
longitudinal momentum correlations between the partons in the same
hadron. In this framework, all the information on the transverse
structure of the proton is captured in the factor $\sigeff$.  The
assumption of factorization between the longitudinal and transverse
components of double parton distributions appears to be not strictly
valid within QCD~\cite{Diehl:2011tt,Gaunt:2011xd,jo}.  However, given
the small values of longitudinal momentum fraction $x$ probed in the
$\jpsi$ production at the LHC, this should be a reasonable
approximation.  For the same reason, the approximation of no
longitudinal momentum correlations may also be justified.

The state--of--the--art theoretical description of direct single
$\jpsi$ production involves the non--relativistic QCD factorization
approach~\cite{Bodwin:1994jh}.  Because of the LHCb ability to trigger
on low $\pt$ muons, down to 1 GeV, it is essential for our analysis to
be able to describe the production of $\jpsi$ with low $\pt$
accurately. Since the fixed--order calculations~\cite{jpsi_corr} fail
to provide such a description in this regime, we resort to modeling
the $\pt$ distribution of the single $\jpsi$.  To do this, we
approximate the matrix element for the inclusive production of a
prompt $\jpsi$, assumed to be dominated by the gluon--gluon channel,
with a crystal ball function of the form
\begin{eqnarray}  
\overline{|M_{gg\to\jpsi+ X}|^2}= \nonumber \hspace{5cm}\\
\left\{
  \begin{array}{ll}
    K\textrm{exp}(-\kappa\frac{\pt^2}{m^2_{\jpsi}}) & \pt \le \mpt\\
    K\textrm{exp}(-\kappa\frac{\mpt^2}{m^2_{\jpsi}})(1+\frac{\kappa}{n}\frac{\pt^2-\mpt^2}{m^2_{\jpsi}})^{-n} & \pt > \mpt\,,
  \end{array} \right.
\label{eq:jpsiparam}
\end{eqnarray} 
where $K=\lambda^2\kappa\hat{s}/m^2_{\jpsi}$.  The values of the
coefficients $\kappa$, $\lambda$, $n$ and $\mpt$ are obtained through
a combined fit of $d \sigma / d\pt$ to the LHCb~\cite{Aaij:2011jh},
ATLAS~\cite{ATLAS:2011sp}, CMS~\cite{Khachatryan:2010yr} and
CDF~\cite{Acosta:2004yw} data, using MSTW 2008 NLO PDFs with the
factorization scale $\mu_F= \sqrt{m_{\jpsi}^2 + \pt^2}$ and the
physical mass $m_{\jpsi}=3.096$ GeV.  The resulting fit gives
$\kappa=0.6$ and $\lambda=0.327$ for $n=2$ and $\mpt=4.5$ GeV.  In
particular, $\chi^2 \approx 28$ for the fit to 55 LHCb data points.

The parametrization~(\ref{eq:jpsiparam}) is then used to obtain the
cross section for the production of two $\jpsi$ mesons through
DPS. The corresponding matrix element is implemented into the
framework of \herwig, with parton shower and intrinsic
$\pt$--broadening switched off as the fitted data already account for
these effects.

The value of $\sigeff$ in the factorized approach to DPS, {\it
  cf}. Eq.~(\ref{eq:DPS}), which could be energy-- and process--
dependent, is one of the properties of the DPS that requires a more
precise experimental measurement.  In the calculations we use
$\sigeff=14.5\,\mb$, a value obtained by the CDF $\gamma+3j$ study
\cite{CDF}.  As clearly seen in Fig.~\ref{fig:res2M}, double $\jpsi$
production at LHCb offers a promising opportunity to probe the DPS
component. In the following, we propose a method to separate the DPS
and SPS contributions.

\begin{figure}
  \includegraphics[scale=\scale]{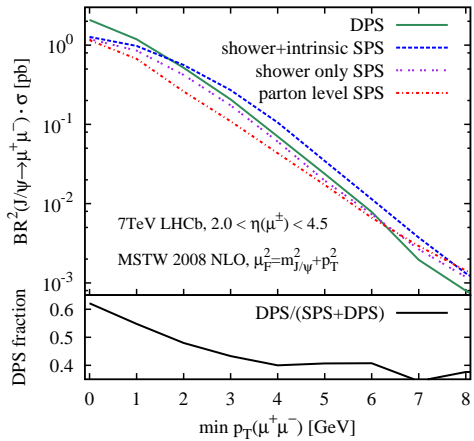}
  \caption{DPS and SPS cross sections for $pp \to \jpsi\jpsi +X$, as a
    function of minimum $\pt$ of $\jpsi$, constructed from $\mup\mum$
    (upper panel), and the fraction of DPS events using the ``shower +
    intrinsic'' SPS calculation (lower panel).  Basic selection cuts
    applied.
    \label{fig:sig_vs_VecPt}}
\end{figure}

{\bf Measurement of DPS at LHCb.} The majority of kinematic variables
used in the literature to distinguish a DPS signal from the SPS
background, where all four final states originate from a single
parton--parton hard scattering, are based on the idea of pair--wise
balancing.  For DPS, the two final state particles from the same hard
scatter will balance against each other on the plane transverse to the
collision axis, resulting in equal but opposite transverse momenta.
The balancing is exact at leading order, but, as we will see in the
following, radiation effects can have significant impact and reduce
the effectiveness of these variables.

In our simulation, we consider events with four muons, each required
to have $\pt>1$ GeV, and to lie within pseudorapidity range $2< \eta
< 4.5$.  Out of the two combinatoric ways to form two $\jpsi$
candidates, the combination with invariant mass closest to the
physical $\jpsi$ mass is chosen. In the rest of this Letter we refer
to this set of cuts as the `basic cuts'.  Once the muon candidates
fulfill these cuts, 100\% reconstruction and detection efficiency is
assumed.  The same parameter values and PDFs as in the calculation of
Fig.~\ref{fig:res2M} are used.

Fig.~\ref{fig:sig_vs_VecPt} shows the DPS and SPS cross sections as a
function of the minimum $\pt$ of a muon pair, after applying the basic
cuts.  In the upper panel, the effects of the parton shower and the
intrinsic $\pt$--broadening on the SPS predictions is shown.  These
effects are significant, due to the low invariant mass of the
two--$\jpsi$ system.  In the lower panel, we see that the DPS fraction
increases as minimum $\pt$ decreases.  Conversely, a cleaner SPS
sample might be obtained by imposing a cut on the minimum $\pt$ of the
$\jpsi$.

\begin{figure}
  \includegraphics[scale=\scale]{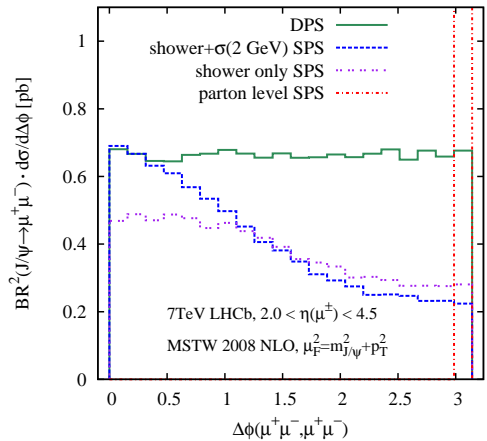}
  \caption{Angular separation $\Delta\phi$ on the plane transverse to
    the beam axis between the two $\jpsi$'s, reconstructed from
    $\mup\mum$ pairs, for DPS and SPS contributions to $pp \to
    \jpsi\jpsi+X $.  Basic selection cuts
    applied. \label{fig:dPhi_vv}}
\end{figure}

The impact of the ISR and the intrinsic $\pt$--broadening is also
demonstrated in Fig.\,\ref{fig:dPhi_vv}, where we show the
distribution of $\dphi\equiv\dphi(\mup\mum,\mup\mum)$, the azimuthal
angular separation between the two reconstructed $\jpsi$'s. As
expected, the signal distribution is flat, a reflection of the
independent scattering hypothesis.  For the background, while
$\dphi=\pi$ at the parton level, the distribution is heavily distorted
in the presence of ISR and $\pt$--broadening.  In particular, the ISR
leads to distributions that are flat or even peaked towards
$\dphi=0$. We conclude that variables based on pair--wise balancing
might not be the best tools to distinguish between DPS and SPS in this
particular analysis.

\begin{table}[t]
  \centering
  \begin{tabular}{|c|r@{.}l|r@{.}l|}
    \hline
    \multicolumn{5}{|c|}{$BR^2 \times$ cross sections [pb] at 7 TeV LHCb}\\
    \hline
      & \multicolumn{2}{|c|}{\;\;\;\;\;\;DPS\;\;\;\;\;\;} &\multicolumn{2}{|c|}{SPS} \\
    \hline
    basic cuts & \;\;\;\;\;\;2&18 & \;\;\;\;\;\;1&27\;\;\;\;\;\;\\
    \hline
    \;basic cuts $+ |\drap|>1.0$\; & 0&48  & 0&07\\
    basic cuts $+ |\drap|>1.5$ & 0&11  & 0&005\\
    basic cuts $+ |\drap|>2.0$ & 0&007 & $<$\;0&001\\
    \hline
  \end{tabular}
  \caption{DPS and SPS cross sections, including branching ratios into
    muons.  The first row shows the cross section after basic
    selection cuts, while the rest shows the impact of an additional
    $\drapmin$ cut.  For clarity, only the SPS result with parton
    shower and intrinsic $\pt$--broadening turned on is shown.}
  \label{tab:xsec_cuts}
\end{table}

However, as we now demonstrate, it is possible to use correlations
along the \emph{longitudinal} direction between the two $\jpsi$ mesons
to extract the DPS signal.  The idea relies on the observation that in
order to minimize the invariant mass of the $\jpsi$ pair, the SPS
background should on average be characterized by a small rapidity
separation ($\drap$).  To see this, note that in a frame where the
$\pt$ of the $\jpsi$--pair is zero,
\begin{eqnarray}
m_{\jpsi \jpsi}&=&2\sqrt{m^2_{\jpsi}+\pt^2}\,\,{\rm cosh}
\left(\frac{\drap}{2} \right),
\end{eqnarray}
hence a small $|\drap|$ is preferred.  However, this constraint does
not apply to the DPS signal, which implies a broader distribution.
The small invariant mass of the system ensures that overall momentum
conservation has negligible impact on the $\rap$, and hence the
$|\drap|$, distributions.

\begin{figure}[!h]
  \includegraphics[scale=\scale]{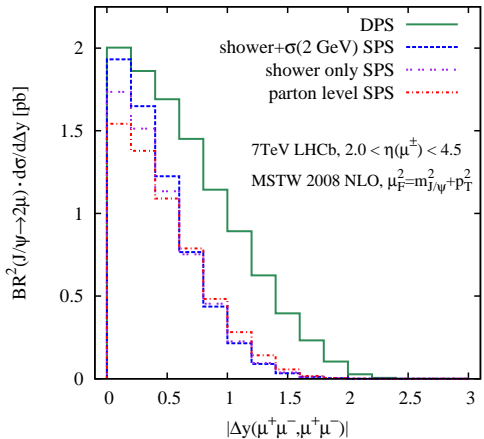}
  \caption{Rapidity separation $|\drap|$ between the two reconstructed
    $\jpsi$'s, with basic selection cuts applied. \label{fig:dEta_vv}}
\end{figure}

\begin{figure}[!h]
  \includegraphics[scale=\scale]{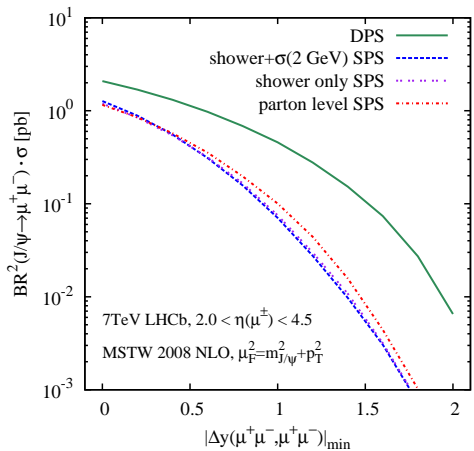}
  \caption{DPS and SPS cross sections for $\jpsi$--pair production
    with basic selection cuts applied, as a function of a lower cut
    $\drapmin$ on rapidity separation. \label{fig:sig_vs_Dy}}
\end{figure}

As shown in Fig.\,\ref{fig:dEta_vv} the difference in $|\drap|$
distributions persists in the laboratory frame.  As expected, the DPS
signal is broader and extends to higher values of $|\drap|$.  The
distributions are also more stable against radiation and intrinsic
$\pt$ effects when compared with the $\dphi$ distributions, making the
predictions more robust.

To extract the DPS signal, we apply a lower cut on the rapidity
separation, $|\drap|> \drapmin$.  The variation of the cross section
with $\drapmin$ is displayed in Fig.\,\ref{fig:sig_vs_Dy}.  Clearly,
the event sample becomes more dominated by DPS contributions for
higher values of $\drapmin$.  A summary of the results is displayed in
Table\,\ref{tab:xsec_cuts}.  In the current (7 TeV) LHC run, an
integrated luminosity of a few $\ifb$ is expected.  By selecting the
four--muon signal sample using the basic cuts, a signal to background
ratio of a few to one may be achieved. Hence we conclude that DPS can
be measured at the LHCb in the four--muon events already at this stage
of LHC running.

{\bf Summary.} Precise measurement of double parton scattering
processes at the LHC is an important step towards understanding
multiple interactions in hadron collisions. The characteristics of the
LHCb detector make it particularly well suited to study DPS in the
production of a $\jpsi$--pair decaying into four muons. We observe
that the first LHCb data on the invariant mass distribution of the
$\jpsi$--pair system might already indicate a significant contribution
from the DPS production mechanism. The studies presented in this
Letter show that it is possible to measure the DPS component in the
four--muon events at the LHCb already in the early stages of the LHC
running, in particular with the help of the proposed rapidity
separation variable $\drap$.

\begin{acknowledgments}
This work has been supported in part by the Helmholtz Alliance
``Physics at the Terascale'', the Isaac Newton Trust and the STFC.  AK
would like to thank the High Energy Physics Group at the Cavendish
Laboratory for hospitality.
\end{acknowledgments}


\end{document}